\documentclass[12pt]{iopart}

\usepackage{bm}
\usepackage{graphicx}
\usepackage{amsfonts}
\usepackage{amssymb}
\usepackage{amsbsy}
\usepackage{array}
\usepackage{iopams}
\usepackage{dsfont}
\usepackage{cite}

\begin{document}

\title[]{Comment on `Energy transfer, entanglement and decoherence in a molecular dimer interacting with a phonon bath'}

\author{James Lim$^{1,2,3}$, Mark Tame$^{4}$, Ki Hyuk Yee$^{1,2}$, \\ Joong-Sung Lee$^{1,2}$ and Jinhyoung Lee$^{1,2}$}

\address{$^1$ Department of Physics, Hanyang University, Seoul 133-791, Korea}
\address{$^2$ Center for Macroscopic Quantum Control, Seoul National University, Seoul 151-742, Korea}
\address{$^3$ Research Institute for Natural Sciences, Hanyang University, Seoul 133-791, Korea}
\address{$^4$ School of Chemistry and Physics, University of KwaZulu-Natal, Durban 4001, South Africa}
\ead{markstame@gmail.com and hyoung@hanyang.ac.kr}

\begin{abstract}
We show that the influence of the shared phonon bath considered in Hossein-Nejad H and Scholes G D 2010 {\it New J. Phys.} {\bf 12} 065045 on the exciton transfer in a two-molecule system can be reproduced by that of an independent bath model.
\end{abstract}

\maketitle

\newcommand{\bra}[1]{\left\langle #1\right|}
\newcommand{\ket}[1]{\left|#1\right\rangle}
\newcommand{\abs}[1]{\left|#1\right|}
\newcommand{\ave}[1]{\left<#1\right>}


\section{Introduction}

In a recent paper~\cite{Hossein-Nejad2010NJP}, Hossein-Nejad and Scholes investigated a two-molecule system coupled to a shared phonon bath, where the molecules are coupled to the same phonon modes. The Hamiltonian of this shared bath model is different from that of an independent bath model where each molecule is coupled to an independent phonon bath. The two bath models are mathematically different and describe different physical situations. However, the authors investigated the time evolution of the reduced electronic state of the molecules, which describes exciton transfer dynamics, and mentioned that the influence of the shared bath model on exciton transfer cannot be reproduced with an independent bath model.

Here we show that an independent bath model can reproduce the same exciton transfer dynamics as obtained by the shared bath model considered by the authors. This means that if one considers only the reduced electronic state of the molecules by tracing out the phonon degrees of freedom, a distinction between the shared and independent bath models is not possible. This comment therefore suggests that in order to distinguish the shared bath model from an independent bath model, one needs to consider something beyond the exciton transfer dynamics, such as the correlations between exciton and phonons.

The comment is organized as follows. In Section 2, we prove the equivalence between the shared bath model considered in Ref.~\cite{Hossein-Nejad2010NJP} and an independent bath model. Here we call two bath models equivalent if they lead to the same exciton transfer dynamics. For simplicity, we prove the equivalence in the presence of a few phonon modes and then generalize the proof to the case of many phonon modes as considered by the authors. In Section 3, we show that a transformation technique employed in Section 2 can provide a new insight into a different type of shared phonon bath considered in Ref.~\cite{Ishizaki2010NJP}.

\section{Anti-correlated Shared Bath}
\subsection{Anti-correlated shared phonon mode}

We investigate single-exciton transfer with the Hamiltonian of the total system in the single-exciton manifold
\begin{equation}
	\hat{H}=\hat{H}_{e}+\hat{H}_{\rm ph}+g(\ket{1}\bra{1}-\ket{2}\bra{2})\otimes(\hat{b}^{\dagger}+\hat{b}),
	\label{eq:Hossein-Nejad2010NJP_simplified}
\end{equation}
where $\ket{j}$ represents a single-exciton state with only molecule $j$ excited and the other molecule is in its ground state. Here, $\hat{H}_{e}$ is the Hamiltonian of the molecules and $\hat{H}_{\rm ph}=\hbar\omega\hat{b}^{\dagger}\hat{b}$ is the Hamiltonian of a shared phonon mode with $\hat{b}^{\dagger}$ and $\hat{b}$ denoting its creation and annihilation operators. The state of the total system at initial time $t=0$ is taken to be in a product form of $\hat{\rho}(0)=\hat{\rho}_{e}(0)\otimes e^{-\beta \hat{H}_{\rm ph}}/{\rm Tr}[e^{-\beta \hat{H}_{\rm ph}}]$ where $\hat{\rho}_{e}(0)$ is the initial single-exciton state and $\beta$ is the inverse temperature of the shared phonon mode. The exciton transfer dynamics are described by the reduced electronic state of the molecules $\hat{\rho}_{e}(t)={\rm Tr_{ph}}[\hat{\rho}(t)]$ where $\hat{\rho}(t)$ is the state of the total system at time $t$ and ${\rm Tr_{ph}}$ is the partial trace over the phonon degrees of freedom.

To demonstrate that the influence of the shared phonon mode modeled by Eq.~\eref{eq:Hossein-Nejad2010NJP_simplified} on the reduced electronic state of molecules $\hat{\rho}_{e}(t)$ is equivalent to that of independent phonon modes, we consider a two-molecule system where molecule $j$ is coupled to an independent phonon mode associated with $\hat{c}_j$
\begin{equation}
	\hat{\cal H}=\hat{H}_{e}+\hat{\cal H}_{\rm ph}+\hbar\sqrt{2}g\ket{1}\bra{1}\otimes(\hat{c}_{1}^{\dagger}+\hat{c}_{1})+\hbar\sqrt{2}g\ket{2}\bra{2}\otimes(\hat{c}_{2}^{\dagger}+\hat{c}_{2}),
	\label{eq:local_model_1}
\end{equation}
where $\hat{\cal H}_{\rm ph}=\hbar\omega\hat{c}_{1}^{\dagger}\hat{c}_{1}+\hbar\omega\hat{c}_{2}^{\dagger}\hat{c}_{2}$ is the Hamiltonian of the phonon modes, and the initial state of the total system is given by $\hat{\rho}(0)=\hat{\rho}_{e}(0)\otimes e^{-\beta \hat{\cal H}_{\rm ph}}/{\rm Tr}[e^{-\beta \hat{\cal H}_{\rm ph}}]$. We consider a unitary transformation (basis change) of the annihilation operators of the phonon modes
\begin{equation}
	\pmatrix{
		\hat{c}_{1}\cr
		\hat{c}_{2}\cr}
	=\frac{1}{\sqrt{2}}
	\pmatrix{
		1 & 1\cr
		-1 & 1\cr}
	\pmatrix{
		\hat{b}\cr
		\hat{B}\cr}
	=U
	\pmatrix{
		\hat{b}\cr
		\hat{B}\cr}.
\end{equation}
Due to unitarity, $UU^{\dagger}=U^{\dagger}U=I$, and the bosonic commutation relations $[\hat{c}_{1},\hat{c}_{1}^{\dagger}]=1$, $[\hat{c}_{2},\hat{c}_{2}^{\dagger}]=1$, $[\hat{c}_{1},\hat{c}_{2}]=[\hat{c}_{1},\hat{c}_{2}^{\dagger}]=0$, the unitary-transformed operators $\hat{b}$ and $\hat{B}$ satisfy the bosonic commutation relations $[\hat{b},\hat{b}^{\dagger}]=1$, $[\hat{B},\hat{B}^{\dagger}]=1$, $[\hat{b},\hat{B}]=[\hat{b},\hat{B}^{\dagger}]=0$, with $\hat{b}^{\dagger}\hat{b}+\hat{B}^{\dagger}\hat{B}=\hat{c}_{1}^{\dagger}\hat{c}_{1}+\hat{c}_{2}^{\dagger}\hat{c}_{2}$. Then the Hamiltonian of the total system $\hat{\cal H}$ can be rewritten as
\begin{eqnarray}
	\hat{\cal H}=\hat{H}_{e}+\hat{\cal H}_{\rm ph}&+\hbar g(\ket{1}\bra{1}-\ket{2}\bra{2})\otimes(\hat{b}^{\dagger}+\hat{b})\nonumber\\
	&+\hbar g(\ket{1}\bra{1}+\ket{2}\bra{2})\otimes(\hat{B}^{\dagger}+\hat{B}),
\end{eqnarray}
where $\hat{\cal H}_{\rm ph}=\hbar\omega\hat{b}^{\dagger}\hat{b}+\hbar\omega\hat{B}^{\dagger}\hat{B}$. Note that the phonon mode associated with $\hat{B}$ does not influence the exciton transfer. This is because: (i) the reduced electronic state $\hat{\rho}_{e}(t)={\rm Tr_{ph}}[\hat{\rho}(t)]$ is in the single-exciton subspace for all time $t$, {\it i.e.} $\hat{\rho}_{e}(t)=\rho_{11}(t)\ket{1}\bra{1}+\rho_{22}(t)\ket{2}\bra{2}+\rho_{12}(t)\ket{1}\bra{2}+\rho_{21}(t)\ket{2}\bra{1}$, and (ii) the initial state of the phonon mode associated with $\hat{B}$ is decoupled from the rest of the total system at time $t=0$. Then the interaction Hamiltonian $\hbar g(\ket{1}\bra{1}+\ket{2}\bra{2})\otimes(\hat{B}^{\dagger}+\hat{B})$ leads to the decoupled dynamics of the phonon associated with $\hat{B}$ and the other degrees of freedom. In this case, the dynamics of the reduced electronic state $\hat{\rho}_{e}(t)$, {\it i.e.} the exciton transfer dynamics, is not influenced even if the phonon mode associated with $\hat{B}$ is removed from $\hat{\cal H}$ such that
\begin{eqnarray}
	\hat{\cal H}'=\hat{H}_{e}+\hat{\cal H}'_{\rm ph}+\hbar g(\ket{1}\bra{1}-\ket{2}\bra{2})\otimes(\hat{b}^{\dagger}+\hat{b}),
\end{eqnarray}
where $\hat{\cal H}'_{\rm ph}=\hbar\omega\hat{b}^{\dagger}\hat{b}$, and the initial state of the total system is given by $\hat{\rho}'(0)=\hat{\rho}_{e}(0)\otimes e^{-\beta \hat{\cal H}'_{\rm ph}}/{\rm Tr}[e^{-\beta \hat{\cal H}'_{\rm ph}}]$, leading to $\hat{\rho}_{e}(t)={\rm Tr_{ph}}[\hat{\rho}(t)]={\rm Tr_{ph}}[\hat{\rho}'(t)]$. This implies that the influence of the shared phonon mode modeled by Eq.~(\ref{eq:Hossein-Nejad2010NJP_simplified}) on exciton transfer is equivalent to that of the independent phonon modes associated with $\hat{c}_{1}$ and $\hat{c}_{2}$ modeled by Eq.~(\ref{eq:local_model_1}).

\subsection{Anti-correlated shared phonon bath}

We now generalize the proof of the equivalence to the case of many phonon modes and show that the shared bath model considered in Ref.~\cite{Hossein-Nejad2010NJP} is equivalent to an independent bath model. In Ref.~\cite{Hossein-Nejad2010NJP}, single-exciton transfer was investigated with the Hamiltonian of the total system in the single-exciton manifold
\begin{equation}
	\hat{H}=\hat{H}_{e}+\hat{H}_{\rm ph}+\sum_{\xi}\hbar g_{\xi}(\ket{1}\bra{1}-\ket{2}\bra{2})\otimes(\hat{b}^{\dagger}_{\xi}+\hat{b}_{\xi}),
	\label{eq:Hossein-Nejad2010NJP}
\end{equation}
where $\hat{H}_{\rm ph}=\sum_{\xi}\hbar\omega_{\xi}\hat{b}_{\xi}^{\dagger}\hat{b}_{\xi}$ is the Hamiltonian of the phonon bath with $\hat{b}_{\xi}^{\dagger}$ and $\hat{b}_{\xi}$ denoting creation and annihilation operators of a shared phonon mode $\xi$. The state of the total system at initial time $t=0$ was taken to be in a product form of $\hat{\rho}(0)=\hat{\rho}_{e}(0)\otimes e^{-\beta \hat{H}_{\rm ph}}/{\rm Tr}[e^{-\beta \hat{H}_{\rm ph}}]$ where $\hat{\rho}_{e}(0)$ is the initial single-exciton state and $\beta$ is the inverse temperature of the phonon bath. The exciton transfer dynamics were described by the reduced electronic state of molecules $\hat{\rho}_{e}(t)={\rm Tr_{ph}}[\hat{\rho}(t)]$.

To demonstrate that the shared phonon bath modeled by Eq.~\eref{eq:Hossein-Nejad2010NJP} is equivalent to an independent bath model, we start with the Hamiltonian of the total system in the presence of independent phonon modes
\begin{equation}
	\hat{\cal H}=\hat{H}_{e}+\hat{\cal H}_{\rm ph}+\sum_{j=1}^{2}\sum_{\xi}\hbar \sqrt{2}g_{\xi}\ket{j}\bra{j}\otimes(\hat{c}_{j\xi}^{\dagger}+\hat{c}_{j\xi}),
	\label{eq:local_model_2}
\end{equation}
where $\hat{\cal H}_{\rm ph}=\sum_{j=1}^{2}\sum_{\xi}\hbar\omega_{\xi}\hat{c}_{j\xi}^{\dagger}\hat{c}_{j\xi}$. We now consider a unitary transformation of the annihilation operators of the phonon modes
\begin{equation}
	\pmatrix{
		\hat{c}_{1\xi}\cr
		\hat{c}_{2\xi}\cr}
	=\frac{1}{\sqrt{2}}
	\pmatrix{
		1 & 1\cr
		-1 & 1\cr}
	\pmatrix{
		\hat{b}_{\xi}\cr
		\hat{B}_{\xi}\cr}
	=U
	\pmatrix{
		\hat{b}_{\xi}\cr
		\hat{B}_{\xi}\cr}.
\end{equation}
Due to unitarity, $UU^{\dagger}=U^{\dagger}U=I$, and the bosonic commutation relations $[\hat{c}_{1\xi},\hat{c}_{1\xi}^{\dagger}]=1$, $[\hat{c}_{2\xi},\hat{c}_{2\xi}^{\dagger}]=1$, $[\hat{c}_{1\xi},\hat{c}_{2\xi}]=[\hat{c}_{1\xi},\hat{c}_{2\xi}^{\dagger}]=0$, the unitary-transformed operators $\hat{b}_{\xi}$ and $\hat{B}_{\xi}$ satisfy the bosonic commutation relations $[\hat{b}_{\xi},\hat{b}_{\xi}^{\dagger}]=1$, $[\hat{B}_{\xi},\hat{B}_{\xi}^{\dagger}]=1$, $[\hat{b}_{\xi},\hat{B}_{\xi}]=[\hat{b}_{\xi},\hat{B}_{\xi}^{\dagger}]=0$, with $\hat{b}_{\xi}^{\dagger}\hat{b}_{\xi}+\hat{B}_{\xi}^{\dagger}\hat{B}_{\xi}=\hat{c}_{1\xi}^{\dagger}\hat{c}_{1\xi}+\hat{c}_{2\xi}^{\dagger}\hat{c}_{2\xi}$. Then the Hamiltonian of the total system $\hat{\cal H}$ can be rewritten as
\begin{eqnarray}
	\hat{\cal H}=\hat{H}_{e}+\hat{\cal H}_{\rm ph}&+\sum_{\xi}\hbar g_{\xi}(\ket{1}\bra{1}-\ket{2}\bra{2})\otimes(\hat{b}_{\xi}^{\dagger}+\hat{b}_{\xi})\nonumber\\
	&+\sum_{\xi}\hbar g_{\xi}(\ket{1}\bra{1}+\ket{2}\bra{2})\otimes(\hat{B}_{\xi}^{\dagger}+\hat{B}_{\xi}),
\end{eqnarray}
where $\hat{\cal H}_{\rm ph}=\sum_{\xi}\hbar\omega_{\xi}\hat{b}_{\xi}^{\dagger}\hat{b}_{\xi}+\sum_{\xi}\hbar\omega_{\xi}\hat{B}_{\xi}^{\dagger}\hat{B}_{\xi}$. Similar to the case of a few phonon modes, the phonon modes associated with $\hat{B}_{\xi}$ do not influence the exciton transfer dynamics even if they are removed from $\hat{\cal H}$. This implies that the influence of the shared phonon modes $\hat{b}_{\xi}$ modeled by Eq.~(\ref{eq:Hossein-Nejad2010NJP}) on exciton transfer is equivalent to that of the independent phonon modes associated with $\hat{c}_{1\xi}$ and $\hat{c}_{2\xi}$ modeled by Eq.~(\ref{eq:local_model_2}). This shows that the shared bath model considered in Ref.~\cite{Hossein-Nejad2010NJP} is equivalent to an independent bath model.


\section{Positive and Negative Correlations}

The transformation technique employed in the previous section can provide a new insight into a different type of shared phonon bath. In Ref.~\cite{Ishizaki2010NJP}, the Hamiltonian of the total system in the single-exciton manifold was modeled by
\begin{eqnarray}
	\hat{H}=\hat{H}_{e}+\hat{H}_{\rm ph}&+\sum_{\xi}\hbar g_{\xi}(\ket{1}\bra{1}+\alpha\ket{2}\bra{2})\otimes(\hat{b}_{1\xi}^{\dagger}+\hat{b}_{1\xi})\nonumber\\
	&+\sum_{\xi}\hbar g_{\xi}(\ket{2}\bra{2}+\alpha\ket{1}\bra{1})\otimes(\hat{b}_{2\xi}^{\dagger}+\hat{b}_{2\xi}),
	\label{eq:Ishizaki2010NJP}
\end{eqnarray}
with $\alpha=0$ corresponding to an independent bath model, $0<\alpha\le 1$ and $\alpha<0$ were called positive and negative correlations respectively. Here, $\hat{H}_{e}$ is the Hamiltonian of the molecules and $\hat{H}_{\rm ph}=\sum_{j=1}^{2}\sum_{\xi}\hbar\omega_{\xi}\hat{b}_{j\xi}^{\dagger}\hat{b}_{j\xi}$ is the Hamiltonian of the phonon modes. The state of the total system at initial time $t=0$ was taken to be in a product form of $\hat{\rho}(0)=\hat{\rho}_{e}(0)\otimes e^{-\beta \hat{H}_{\rm ph}}/{\rm Tr}[e^{-\beta \hat{H}_{\rm ph}}]$ where $\hat{\rho}_{e}(0)$ is the initial single-exciton state and $\beta$ is the inverse temperature of the phonon bath. Similarly to the previous case, we consider a unitary transformation of the annihilation operators of the phonon modes
\begin{equation}
	\pmatrix{
		\hat{b}_{1\xi}\cr
		\hat{b}_{2\xi}\cr}
	=\frac{1}{\sqrt{2}}
	\pmatrix{
		1 & 1\cr
		-1 & 1\cr}
	\pmatrix{
		\hat{c}_{\xi}\cr
		\hat{C}_{\xi}\cr}
	=U
	\pmatrix{
		\hat{c}_{\xi}\cr
		\hat{C}_{\xi}\cr}.
\end{equation}
Due to unitarity, $UU^{\dagger}=U^{\dagger}U=I$, and the bosonic commutation relations $[\hat{b}_{1\xi},\hat{b}_{1\xi}^{\dagger}]=1$, $[\hat{b}_{2\xi},\hat{b}_{2\xi}^{\dagger}]=1$, $[\hat{b}_{1\xi},\hat{b}_{2\xi}]=[\hat{b}_{1\xi},\hat{b}_{2\xi}^{\dagger}]=0$, the unitary-transformed operators $\hat{c}_{\xi}$ and $\hat{C}_{\xi}$ satisfy the bosonic commutation relations $[\hat{c}_{\xi},\hat{c}_{\xi}^{\dagger}]=1$, $[\hat{C}_{\xi},\hat{C}_{\xi}^{\dagger}]=1$, $[\hat{c}_{\xi},\hat{C}_{\xi}]=[\hat{c}_{\xi},\hat{C}_{\xi}^{\dagger}]=0$, with $\hat{c}_{\xi}^{\dagger}\hat{c}_{\xi}+\hat{C}_{\xi}^{\dagger}\hat{C}_{\xi}=\hat{b}_{1\xi}^{\dagger}\hat{b}_{1\xi}+\hat{b}_{2\xi}^{\dagger}\hat{b}_{2\xi}$. Then the Hamiltonian of the total system $\hat{H}$ can be rewritten as
\begin{eqnarray}
	\hat{H}=\hat{H}_{e}+\hat{H}_{\rm ph}&+\sum_{\xi}\hbar\frac{g_{\xi}}{\sqrt{2}}(1-\alpha)(\ket{1}\bra{1}-\ket{2}\bra{2})\otimes(\hat{c}_{\xi}^{\dagger}+\hat{c}_{\xi})\nonumber\\
	&+\sum_{\xi}\hbar \frac{g_{\xi}}{\sqrt{2}}(1+\alpha)(\ket{1}\bra{1}+\ket{2}\bra{2})\otimes(\hat{C}_{\xi}^{\dagger}+\hat{C}_{\xi}),
\end{eqnarray}
with $\hat{H}_{\rm ph}=\sum_{\xi}\hbar\omega_{\xi}\hat{c}_{\xi}^{\dagger}\hat{c}_{\xi}+\sum_{\xi}\hbar\omega_{\xi}\hat{C}_{\xi}^{\dagger}\hat{C}_{\xi}$. It is notable that the phonon modes associated with $\hat{C}_{\xi}$ do not influence the single-exciton transfer described by the reduced electronic state of the molecules, $\hat{\rho}_{e}(t)$. This implies that the exciton transfer dynamics are not altered even if we consider only the phonon modes associated with $\hat{c}_{\xi}$ such that the Hamiltonian is reduced to
\begin{equation}
	\hat{H}'=\hat{H}_{e}+\hat{H}'_{\rm ph}+\sum_{\xi}\hbar\frac{g_{\xi}}{\sqrt{2}}(1-\alpha)(\ket{1}\bra{1}-\ket{2}\bra{2})\otimes(\hat{c}_{\xi}^{\dagger}+\hat{c}_{\xi}),
	\label{eq:Ishizaki2010NJPequivalence}
\end{equation}
with $\hat{H}'_{\rm ph}=\sum_{\xi}\hbar\omega_{\xi}\hat{c}_{\xi}^{\dagger}\hat{c}_{\xi}$, and the initial state of the total system is given by $\hat{\rho}'(0)=\hat{\rho}_{e}(0)\otimes e^{-\beta \hat{H}'_{\rm ph}}/{\rm Tr}[e^{-\beta \hat{H}'_{\rm ph}}]$, leading to $\hat{\rho}_{e}(t)={\rm Tr_{ph}}[\hat{\rho}(t)]={\rm Tr_{ph}}[\hat{\rho}'(t)]$. This implies that the influence of the shared phonon modes modeled by Eq.~(\ref{eq:Ishizaki2010NJP}) on the exciton transfer is essentially equivalent to that of the shared phonon modes modeled by Eq.~(\ref{eq:Hossein-Nejad2010NJP}) if the exciton-phonon coupling $g_{\xi}$ in Eq.~(\ref{eq:Hossein-Nejad2010NJP}) is replaced by $g_{\xi}(1-\alpha)/\sqrt{2}$ in Eq.~(\ref{eq:Ishizaki2010NJPequivalence}). Thus, this model is also equivalent to a local phonon bath model with an effective exciton-phonon coupling $g_{\xi}(1-\alpha)/\sqrt{2}$, which depends on the degree of correlation $\alpha$. This implies that the positive correlation with $0<\alpha\le 1$ is equivalent to weakly coupled local baths when compared to the independent bath model with $\alpha=0$, {\it i.e.} $\abs{g_{\xi}(1-\alpha)/\sqrt{2}}<\abs{g_{\xi}/\sqrt{2}}$. On the other hand, the negative correlation with $\alpha<0$ is equivalent to strongly coupled local baths when compared to the independent bath model with $\alpha=0$, {\it i.e.} $\abs{g_{\xi}(1-\alpha)/\sqrt{2}}>\abs{g_{\xi}/\sqrt{2}}$. These results suggest that the positive correlation enhances coherent behavior of the exciton transfer as $\alpha$ is increased from 0 up to 1, while the negative correlation suppresses electronic coherence as $\alpha$ is decreased. This is inline with the simulation results displayed in figure 4 of Ref.~\cite{Ishizaki2010NJP}.

\ack
This work was supported by National Research Foundation of Korea grants funded by the Korean Government (Ministry of Education, Science and Technology; grant numbers 2010-0018295 and 2010-0015059), the UK's Engineering and Physical Sciences Research Council, and the Leverhulme Trust.

\end{document}